\documentclass[12pt]{article}

\usepackage[margin=1in]{geometry}
\usepackage{setspace}
\usepackage[T1]{fontenc}
\usepackage{lmodern}
\usepackage{amsmath,amssymb,amsthm,mathtools}
\usepackage{booktabs}
\usepackage{microtype}
\usepackage[authoryear,round,longnamesfirst]{natbib}
\setcitestyle{aysep={,},yysep={,},notesep={, }}

\makeatletter
\DeclareRobustCommand{\APAand}{\ifNAT@swa\&\else and\fi}
\makeatother
\usepackage[hyperfootnotes=false]{hyperref}
\usepackage{graphicx}
\graphicspath{{./}{figures/}}
\usepackage[section]{placeins}
\usepackage{subcaption}
\usepackage{enumitem}

\hypersetup{
  colorlinks=true,
  linkcolor=black,
  citecolor=black,
  urlcolor=black
}

\newtheorem{theorem}{Theorem}[section]

\newcommand{\E}{\mathbb{E}}
\newcommand{\Pp}{\mathbb{P}}

\newcommand{\Ww}{\mathbb{W}}
\newcommand{\F}{\mathcal{F}}
\newcommand{\G}{\mathcal{G}}
\newcommand{\A}{\mathcal{A}}
\newcommand{\CE}{\mathrm{CE}}
\newcommand{\R}{\mathbb{R}}

\title{Long-Horizon Investment with Delayed Information\\in a Mean-Reverting Market}
\author{Yan Dolinsky\\[0.5em]
  {\small Department of Statistics}\\
  {\small The Hebrew University of Jerusalem, Jerusalem, Israel}}
\date{}

\begin{document}
\doublespacing
\maketitle

\begingroup
\renewcommand{\thefootnote}{}
\footnotetext{\textit{Email address:}
\href{mailto:yan.dolinsky@mail.huji.ac.il}{yan.dolinsky@mail.huji.ac.il}\\
The author is partially supported by the Israel Science Foundation
(grant No.\ 305/25).}
\endgroup

\begin{abstract}
We study long-term investment in an Ornstein–Uhlenbeck diffusion model with exponential utility. We solve the finite-horizon problem explicitly and show that information delay qualitatively alters long-horizon behavior. Under full information, the certainty equivalent generated by mean reversion grows quadratically with the investment horizon. With any fixed positive delay, it grows only linearly, at an asymptotic rate that we compute explicitly. We derive an explicit causal representation of the optimal delayed portfolio and show that projecting the full-information strategy onto delayed information can fail dramatically: the resulting exponential loss becomes infinite for sufficiently long horizons. Numerical examples illustrate the economic impact of information delay.
\end{abstract}


\section{Introduction}

In this paper we study exponential utility maximization in an
Ornstein--Uhlenbeck market with delayed information. The investor observes
the price with a fixed delay, while trading takes place at the current
price. The parameters are known, trading is frictionless, and positions
are unrestricted. Our aim is to understand the effect of the delay on
investment over a long time horizon.

Information delay is relevant in practice because prices may arrive late,
or time may be needed to collect data, estimate a signal, and make a
trading decision. The order is then executed at a price which is not yet
observed. This matters for mean-reversion strategies: the price deviation
used to choose a position may have partly disappeared by the time the
trade is made. Our model isolates this effect from transaction costs and
parameter uncertainty.

Although the price process is Markovian, the control problem with delay
is not Markovian in the observed price. The optimal position depends on
the observation history, not only on the last available price. A dynamic programming approach based only on the last observed price
therefore does not close; a path-dependent formulation would be needed. General
control problems with delayed information are studied by
\citet{SaporitoZhang2019}. Here we use the
Gaussian structure to obtain an explicit solution.

For the full-information problem, \citet{GNR2021} showed that optimal
positions increase with the remaining investment horizon. For every fixed initial price, the certainty equivalent grows quadratically
with the horizon; an initial displacement from the long-run mean adds a linear term. The certainty equivalent is the certain amount
whose utility equals the optimal expected utility. We show
that any fixed positive delay changes this growth from quadratic to
linear. We obtain an explicit finite-horizon value, the long-run growth
rate, and the logarithmic correction. The value separates into the gain
from deterministic trading and the additional value of delayed observations.

The optimal strategy is given by a linear Volterra equation driven by
observed prices and can be computed by a simple numerical recursion.
The value approaches the full-information value as the delay tends to
zero at a fixed horizon, but this convergence is not uniform in the
horizon. An initial price away from the long-run mean contributes the
same linear term under both information structures, so the absolute
certainty-equivalent cost of delay is independent of this displacement.

A natural alternative is to replace the current price in the full-information strategy by its conditional expectation given the delayed observations. We show that this rule can lead to infinite expected exponential loss for sufficiently long horizons and provide an explicit horizon threshold beyond which divergence occurs. Thus, projecting the full-information optimizer onto delayed information does not solve the delayed problem.

Our results complement the long-horizon analysis of \citet{GNR2021}
and connect to the literature on investment under partial information.
\citet{Lakner1995} studies both utility from consumption and utility from
terminal wealth when investment and consumption must be adapted to the
filtration generated by asset prices. His martingale approach identifies
optimal consumption and terminal wealth and establishes the existence
of a financing strategy. \citet{Lakner1998} addresses the further task of
representing the optimal trading strategy explicitly. That paper treats
a broad class of utility functions and includes a model with an
unobserved Gaussian drift. The distinction between characterizing
optimal terminal wealth and constructing its implementing portfolio is
also relevant here, where we derive both an exact value and a causal
representation of the optimizer.

In Lakner's setting the investor observes current prices but cannot
observe all the factors determining returns. In our setting the
parameters and price dynamics are known, while recent prices themselves
are unavailable. Consequently, the gain from trading is not adapted to
the investor's delayed filtration. Filtering the current price is
straightforward, but inserting that filter into the full-information
optimizer does not account for the exponential cost of unobserved gains.
Theorem~\ref{prop:naive} makes this distinction explicit.

Further related work includes \citet{BjorkDavisLanden2010}, who obtain
representations of optimal terminal wealth and portfolios for power,
logarithmic, and exponential utility with unobserved return processes.
\citet{ManiaSantacroce2010} formulate exponential utility maximization
under partial information in terms of observable processes and
characterize the value and optimizer by a backward stochastic
differential equation. These results emphasize that reducing an
information set requires solving the resulting optimization problem,
rather than simply projecting an optimizer from a larger filtration.
Our explicit Gaussian calculation isolates the effect of a fixed
observation delay on the long-horizon value and on the integrability of
a projected strategy.

Related work on superreplication, utility indifference pricing, and
semistatic hedging includes
\citep{DolinskyZouari2020,BankDolinsky2021,DolinskyZuk2025}.
Gaussian utility maximization has been studied in discrete time
\citep{DolinskyZuk2023} and in continuous time with delayed information
\citep{Dolinsky2025}. We apply the continuous-time result and solve the
resulting kernel equations explicitly for the Ornstein--Uhlenbeck model.

The paper is organized as follows. Section~\ref{sec:model} introduces the
model, states the main results, and gives numerical illustrations.
Sections~\ref{sec:exact-proof} and \ref{sec:naive} contain the proofs of
Theorems~\ref{prop:finite} and \ref{prop:naive}, respectively.

\section{Main results and numerical illustrations}
\label{sec:model}

Let $W$ be a standard Brownian motion on a probability space
$(\Omega,\F,\Pp)$. The risky price satisfies
\begin{equation*}
 dr_t=\kappa(\theta-r_t)\,dt+\sigma\,dW_t,\qquad r_0=x,
\end{equation*}
where $\kappa,\sigma>0$ and $x,\theta\in\R$. Let
$\F_t$ be the augmented natural filtration of the price. The cash account
bears no interest. For a fixed delay $H>0$, the investor's information is
$\G_t=\F_{(t-H)^+}$. A strategy specifies the number of units held in the
risky asset, and the strategy class over a horizon $T>0$ is
\[
 \A_H(T)=\left\{\gamma:\ \gamma\text{ is }\G\text{-progressively measurable},
 \quad\int_0^T\gamma_t^2dt<\infty\quad\Pp\text{-a.s.}\right\}.
\]
Trading takes place at the current price. Starting from zero capital, the
terminal gain is $V_T^\gamma=\int_0^T\gamma_t\,dr_t$.
This integral is taken in the full price filtration $\F$; a
$\G$-progressively measurable strategy is also $\F$-progressively
measurable. 

We use exponential utility $U(z)=-e^{-z}$. Taking absolute risk aversion
$\alpha=1$ entails no loss of generality: for any $\alpha>0$, the map
$\gamma\mapsto\alpha\gamma$ is a bijection of $\A_H(T)$ and
$\alpha V_T^\gamma=V_T^{\alpha\gamma}$. Hence the minimum exponential
loss is independent of $\alpha$. The optimal position and monetary
certainty equivalent for general $\alpha$ are obtained by dividing the
corresponding quantities for $\alpha=1$ by $\alpha$. The same argument applies to
the full-information admissible class specified below.
This normalization uses the absence of fixed position constraints.

The optimization problem is
\begin{equation}
 V_H(T)=\inf_{\gamma\in\A_H(T)}
 \E\exp\left(-\int_0^T\gamma_t\,dr_t\right).
 \label{eq:value-intro}
\end{equation}
We write
$\CE_H(T)=-\log V_H(T)$ for the monetary certainty equivalent; thus
$U(\CE_H(T))$ equals the maximal expected utility. Set
\[
 \nu=\frac{\kappa^2(\theta-x)^2}{2\sigma^2},\qquad
 \beta_H=\frac{\kappa}{2}(1-e^{-2\kappa H}),\qquad
 \lambda_H=\frac{\kappa}{e^{2\kappa H}-1},\qquad
 p_H=\frac{\lambda_H}{\beta_H}.
\]

\begin{theorem}
\label{prop:finite}\label{thm:longrun}
For every $T,H>0$, the infimum in \eqref{eq:value-intro} is attained and
\begin{equation}
 \CE_H(T)=\nu T+
 \lambda_H\left[(T-H)^+-\frac{\log(1+\beta_H(T-H)^+)}{\beta_H}\right].
 \label{eq:exact-ce}
\end{equation}
The optimal strategy is unique up to $dt\otimes\Pp$-null sets and has the continuous version 
\begin{equation}
 \widehat\gamma_t=
 \frac{\kappa[1+\kappa(T-t)](\theta-x)-Y_t}{\sigma^2},
 \label{eq:optimal-portfolio}
\end{equation}
where $Y_t=0$ for $t\le H\wedge T$ and, for $H<t\le T$, $Y$ is the
unique continuous, $\G$-adapted solution of
\begin{equation}
 Y_t=C_t\left[r_{t-H}-x+\int_{t-H}^t
 \cosh\bigl(\kappa(H-t+s)\bigr)Y_s\,ds\right],\qquad
 C_t=\frac{\kappa[1+\kappa(T-t)]}
 {e^{\kappa H}+\kappa(T-t)\sinh(\kappa H)}.
 \label{eq:volterra}
\end{equation}
For fixed $H>0$, as $T\to\infty$,
\begin{equation}
 \CE_H(T)=(\nu+\lambda_H)T-p_H\log T+O(1).
 \label{eq:long-ce}
\end{equation}
\end{theorem}

Write $\Psi_H(T)=\CE_H(T)-\nu T$. The term $\nu T$ is attainable by
deterministic trading; $\Psi_H(T)$ is the additional value of delayed
observations. If $H\ge T$, no new observation arrives before the horizon
and $\Psi_H(T)=0$. The strategy in \eqref{eq:optimal-portfolio} can be
computed from observed prices by solving \eqref{eq:volterra} forward.

For the full-information benchmark, we apply
\citet[Section~3, equations~(20)--(21)]{GNR2021}, with zero average
price growth and unit risk aversion:
\begin{equation*}
 \gamma_t^{\rm FI}=\frac{\kappa[1+\kappa(T-t)]}{\sigma^2}(\theta-r_t),
 \qquad \CE_0(T)=\nu T+\frac{\kappa^2T^2}{4}.
\end{equation*}

\paragraph{An initial price away from the mean.}
The case $x=\theta$ isolates the value of learning future price
fluctuations, but is not needed for the growth comparison. For every fixed
$x$, the two values are
\[
 \CE_H(T)=\frac{\kappa^2(\theta-x)^2}{2\sigma^2}T+\Psi_H(T),
 \qquad
 \CE_0(T)=\frac{\kappa^2(\theta-x)^2}{2\sigma^2}T
             +\frac{\kappa^2T^2}{4}.
\]
Thus $x\ne\theta$ raises the delayed long-run rate from $\lambda_H$ to
$\nu+\lambda_H$, while the full-information value still grows
quadratically. Equal initial displacements above and below $\theta$ give
the same value: their magnitudes determine the opportunity, and their
signs determine the initial trading direction.

The term $\nu T$ is already attainable without observing any new price.
Indeed, write $\delta=\theta-x$ and use the deterministic position
$\gamma_t^{\rm det}=\kappa\delta[1+\kappa(T-t)]/\sigma^2$. Integration
by parts in the price equation gives
\[
 \int_0^T\gamma_t^{\rm det}\,dr_t
 =\frac{\kappa^2\delta^2T}{\sigma^2}
  +\frac{\kappa\delta}{\sigma}W_T.
\]
The gain is Gaussian, so its certainty equivalent is its mean minus
one half of its variance, namely $\nu T$. Although the expected initial deviation decays, the optimal deterministic
position also depends on the horizon and is not a fixed buy-and-hold
position. Its Gaussian gain identity explains the linear value.

For $t\le H\wedge T$, the optimal delayed strategy is exactly
$\gamma_t^{\rm det}$: it starts long when $x<\theta$ and short when
$x>\theta$. After time $H$, the history-dependent term $Y$ adjusts this
position as observations arrive. The equation for $Y$ uses $r_{t-H}-x$,
whose mean is $(\theta-x)(1-e^{-\kappa(t-H)})$; consequently $Y$ is
generally not centered when $x\ne\theta$. The additive value decomposition
does not assert that the random adjustment has zero mean or contributes
an independent gain.

The certainty-equivalent cost of delay is
\[
 \CE_0(T)-\CE_H(T)=\frac{\kappa^2T^2}{4}-\Psi_H(T),
\]
which is independent of the initial displacement and of $\sigma$.
The proportional cost does depend on the displacement: a larger $\nu T$
makes the same absolute loss a smaller fraction of the full-information
value at a fixed horizon. Nevertheless, for every fixed $x$ and $H>0$,
$\CE_H(T)/\CE_0(T)\to0$ as $T\to\infty$.
If $H\ge T$, the information contribution vanishes and the entire value
is $\nu T$; it is zero only in the centered case.

Since the full-information optimizer is affine in the price, its
conditional projection amounts to replacing the current price by its
conditional mean. This gives
\begin{equation}
 \gamma_t^N=\E[\gamma_t^{\rm FI}\mid\G_t]
 =\frac{\kappa[1+\kappa(T-t)]}{\sigma^2}
 e^{-\kappa\min(t,H)}(\theta-r_{(t-H)^+}).
 \label{eq:naive-formula-main}
\end{equation}
The next result shows that this substitution may lead to infinite
exponential loss.

\begin{theorem}
\label{prop:naive}
Fix $\kappa,H>0$. For every finite $T$, $\gamma^N\in\A_H(T)$ and
$\E\int_0^T(\gamma_t^N)^2dt<\infty$. Nevertheless, for every
$x,\theta\in\R$ and $\sigma>0$,
\begin{equation}
 \E\exp\left(-\int_0^T\gamma_t^N\,dr_t\right)=+\infty,
 \qquad T\ge\overline T_H,
 \label{eq:naive-infinite-main}
\end{equation}
where
\begin{equation}
 \overline T_H=2H+e^{\kappa H}
 \left(\frac{2H}{3\pi}+\frac{3\pi}{2\kappa^2H}\right).
 \label{eq:TH-main}
\end{equation}
\end{theorem}
The bound \eqref{eq:TH-main} is sufficient and need not be the first
divergence horizon.

\subsection{Numerical illustrations}
\label{sec:numerics}

The formulas are easiest to read in units of the mean-reversion speed. Set
\[
  z=\kappa H,\qquad \tau=\kappa T.
\]
The single number $z$ measures how stale the information is relative to the speed at which the signal decays. In the centered case $x=\theta$,
\begin{equation*}
  \frac{1}{\kappa}\lim_{T\to\infty}\frac{\CE_H(T)}{T}
  =\frac{1}{e^{2z}-1}.
\end{equation*}
For $x\ne\theta$, add $\nu/\kappa=\kappa(\theta-x)^2/(2\sigma^2)$ to this normalized total rate. All tables and plots of $\Psi_H$ therefore also apply away from the mean; the total certainty equivalent is obtained by adding $\nu T$. This normalization makes the comparative statics independent of the calendar unit.

\paragraph{The effect of delay.}

Table~\ref{tab:latency} expresses the delay both in mean-reversion units $z=\kappa H$ and as a fraction of the half-life $h_{1/2}=\log(2)/\kappa$. The last column is the normalized long-run information value in the preceding normalized-rate formula.

\begin{table}[ht]
\centering
\caption{Long-run information value as a function of latency.}
\label{tab:latency}
\begin{tabular}{ccc}
\toprule
$\kappa H$ & $H/h_{1/2}$ & $\lambda_H/\kappa$ \\
\midrule
0.05 & 0.072 & 9.508 \\
0.10 & 0.144 & 4.517 \\
0.25 & 0.361 & 1.541 \\
0.50 & 0.721 & 0.582 \\
1.00 & 1.443 & 0.157 \\
2.00 & 2.885 & 0.019 \\
\bottomrule
\end{tabular}
\end{table}

\begin{figure}[htbp]
\centering
\includegraphics[width=0.78\textwidth]{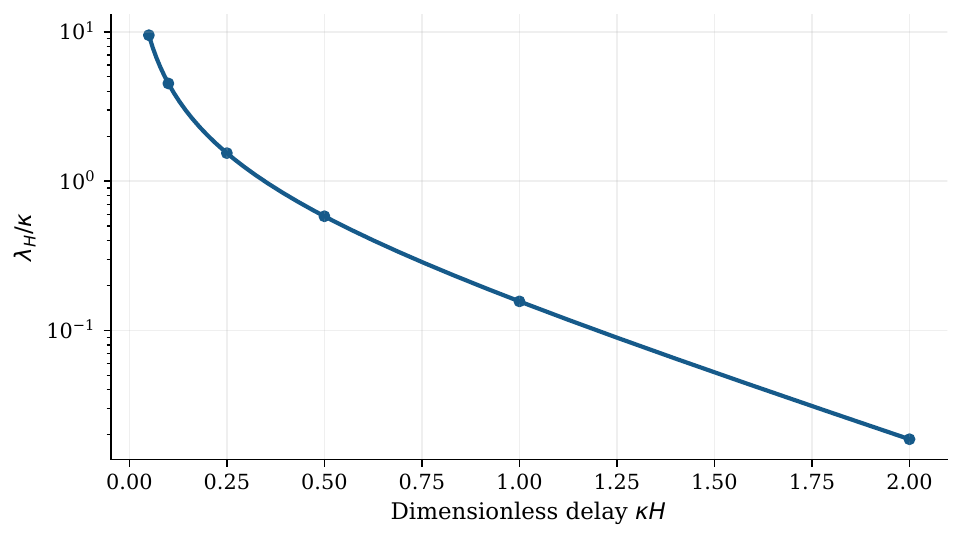}
\caption{Long-run information value per unit time. The curve is evaluated
from the exact formula; markers correspond to Table~\ref{tab:latency}.
The vertical axis is logarithmic.}
\label{fig:delay-cost}
\end{figure}

For a delay of about $14\%$ of the mean-reversion half-life, the rate
is $4.52\kappa$. At a delay of $1.44$ half-lives it is $0.157\kappa$,
and after almost three half-lives it is $0.019\kappa$. Thus the cost of
delay depends on the time scale of the signal, rather than on the delay
measured in calendar units. Figure~\ref{fig:delay-cost} shows the exact
rate across delays.

\paragraph{Long horizons.}

We next take $\kappa=1$ and $x=\theta$. Then the delayed certainty equivalent is exactly $\Psi_H(T)$ and
\[
  \lambda_H=\frac{1}{e^{2H}-1}.
\]
Table~\ref{tab:rates} reports the exact information value per unit time. The convergence to a constant contrasts with full information, for which $\CE_0(T)/T=T/4$ keeps increasing.

\begin{table}[ht]
\centering
\caption{Exact information certainty equivalent per unit time, $\Psi_H(T)/T$, for $\kappa=1$.}
\label{tab:rates}
\begin{tabular}{ccccc}
\toprule
Delay $H$ & $T=10$ & $T=100$ & $T=1000$ & Long-run rate $\lambda_H$ \\
\midrule
0.25 & 0.664 & 1.301 & 1.500 & 1.541 \\
0.50 & 0.297 & 0.515 & 0.571 & 0.582 \\
1.00 & 0.083 & 0.141 & 0.154 & 0.157 \\
2.00 & 0.009 & 0.017 & 0.018 & 0.019 \\
\bottomrule
\end{tabular}
\end{table}

For example, with $\kappa=\sigma=1$, $H=0.5$, and $T=10$,
the centered delayed and full-information values are approximately $2.975$
and $25$. If $|x-\theta|=1$, the term $\nu T=5$ raises these values
to $7.975$ and $30$, respectively; the cost of delay remains $22.025$.

\begin{figure}[htbp]
\centering
\includegraphics[width=\textwidth]{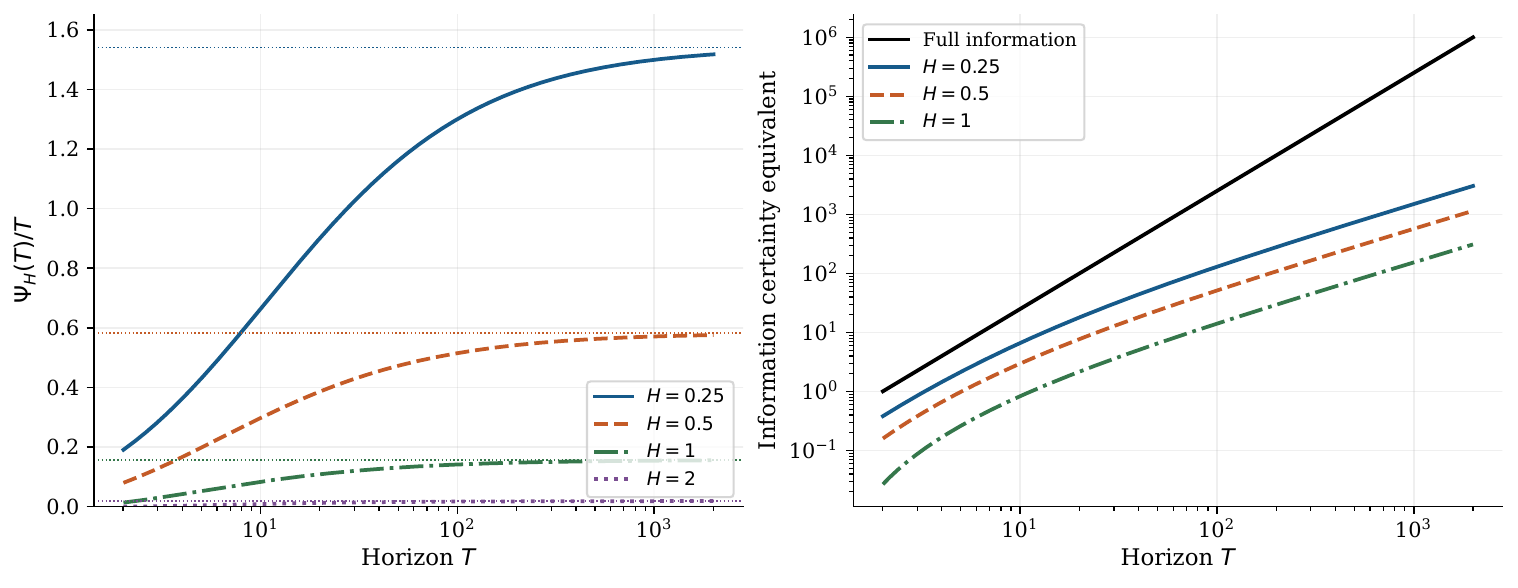}
\caption{Long-horizon value for $\kappa=1$ and $x=\theta$. Left: the
information value per unit time approaches a finite rate; thin dotted
horizontal lines show $\lambda_H$. Right: the full-information value
$T^2/4$ is compared with the exact delayed values. Curves use dense
logarithmic grids rather than interpolation between table entries.}
\label{fig:long-horizon}
\end{figure}

For $H=0.25$, the value per unit time is $1.30$ at $T=100$ and
$1.50$ at $T=1000$, close to the limit $1.54$. Under full information the
corresponding values are $25$ and $250$. The gap therefore increases with
the horizon, as illustrated in Figure~\ref{fig:long-horizon}.

\begin{figure}[htbp]
\centering
\includegraphics[width=\textwidth]{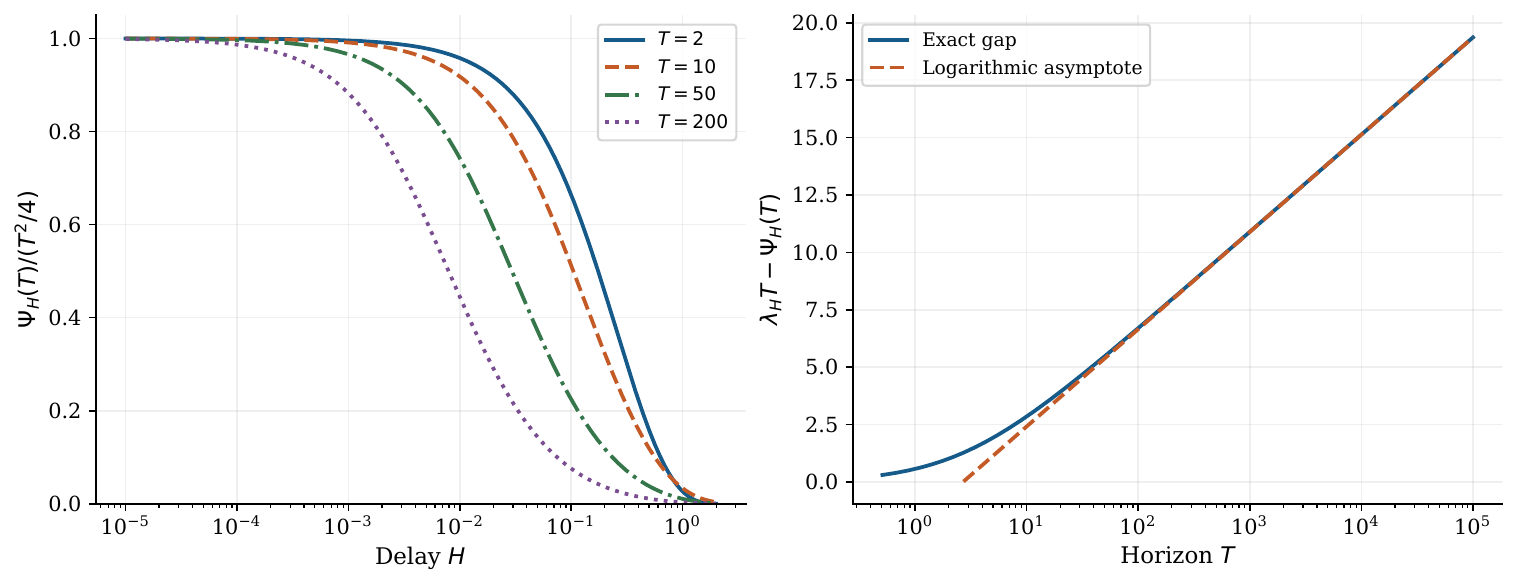}
\caption{Small-delay convergence and the logarithmic correction, with
$\kappa=1$. Left: the fraction of the full-information value retained
at different fixed horizons, evaluated from the exact formula. Right:
the exact gap from the linear asymptote and its logarithmic
approximation for $H=0.5$. The approximation is displayed where positive.}
\label{fig:limits}
\end{figure}

\paragraph{Small and large delays.}
The long-run rate admits the expansions
\[
 \lambda_H=\frac{1}{2H}-\frac{\kappa}{2}+\frac{\kappa^2H}{6}+O(H^3),
 \quad H\downarrow0,\qquad
 \lambda_H\sim\kappa e^{-2\kappa H},\quad H\to\infty.
\]
The information value therefore decays exponentially at large delays.
For a fixed horizon, the value converges to the full-information value as
the delay vanishes. Indeed,
\[
 \Psi_H(T)=\frac{\kappa^2e^{-2\kappa H}}{2}
 \int_0^{(T-H)^+}\frac{q}{1+\beta_Hq}dq
 \longrightarrow\frac{\kappa^2T^2}{4},\qquad H\downarrow0,
\]
by dominated convergence. This differs from the linear growth at every
fixed positive delay: the delay and horizon limits do not commute.
Position constraints, transaction costs, and parameter uncertainty are
not included in this comparison.
Figure~\ref{fig:limits} shows both the fixed-horizon small-delay limit
and the logarithmic correction to linear growth. For $T>H$,
\[
 \lambda_HT-\Psi_H(T)
 =\lambda_HH+p_H\log(1+\beta_H(T-H)),
\]
so its large-$T$ asymptote is
$\lambda_HH+p_H\log(\beta_HT)$.

\begin{figure}[htbp]
\centering
\includegraphics[width=0.80\textwidth]{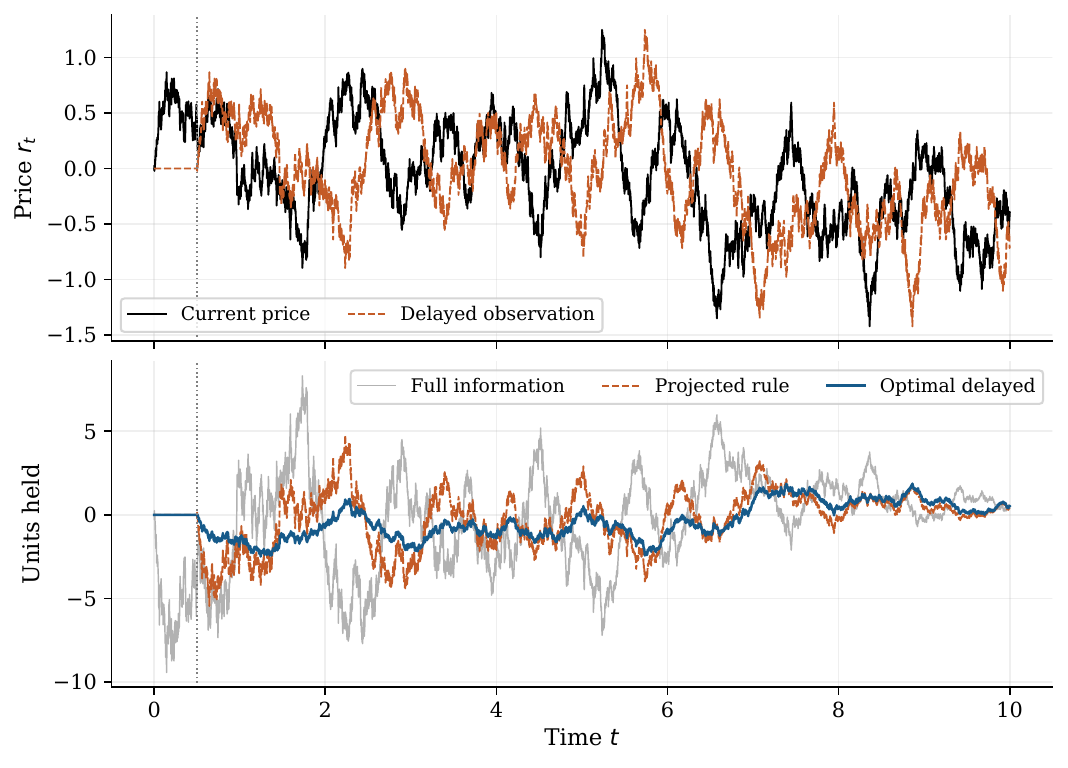}
\caption{A simulated price path and the associated portfolios. Top:
current and delayed prices. Bottom: the full-information optimizer, its
conditional projection, and the optimal delayed strategy. The vertical
dotted line marks the first observation time $H$; in the centered case
both delayed strategies hold zero units before that time.}
\label{fig:portfolios}
\end{figure}

\paragraph{Portfolio paths and the projected rule.}
Figure~\ref{fig:portfolios} compares the three strategies on one simulated
price path, with $\kappa=\sigma=1$, $x=\theta=0$, $T=10$, and $H=0.5$.
The path is generated using exact Ornstein--Uhlenbeck transitions on a
uniform grid with $\Delta=1/1600$ and random seed $20261011$.
The optimal delayed position is computed by the forward recursion in
Section~\ref{sec:portfolio}. For this same path, the maximum absolute
position differences between grids with $\Delta=1/200,1/400,1/800$
and the plotted grid are $0.01551$, $0.00897$, and $0.00244$,
respectively. These are refinement comparisons on a common price path,
rather than bounds on the error relative to the continuous solution.
The path is illustrative: it does not estimate expected utility, and
finite simulated samples cannot establish finiteness of an exponential
moment.

\begin{figure}[htbp]
\centering
\includegraphics[width=0.78\textwidth]{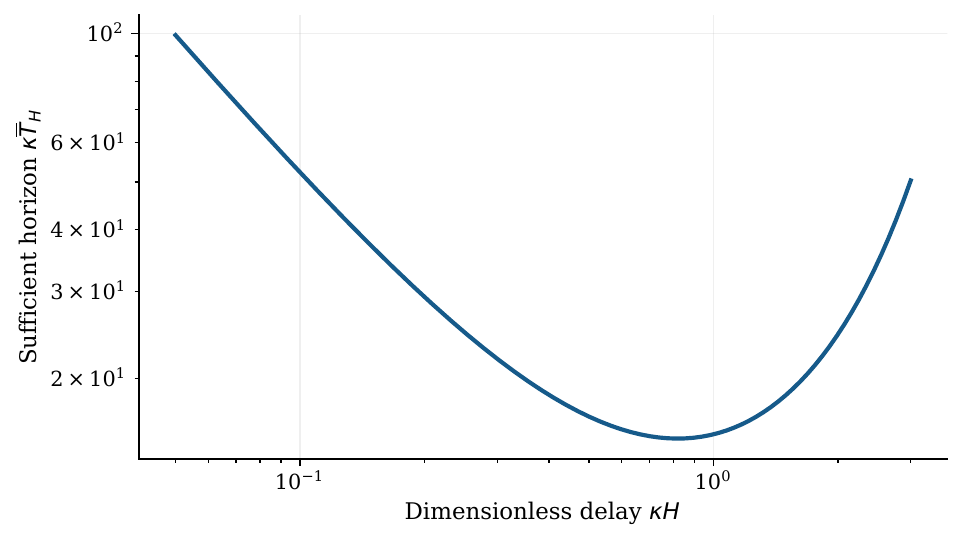}
\caption{The sufficient divergence horizon in Theorem~\ref{prop:naive}.
Above this curve the projected rule has infinite exponential loss.
The curve is an upper bound on the first divergence horizon and makes
no assertion of finite loss below it. Both axes are logarithmic.}
\label{fig:divergence-bound}
\end{figure}

For the divergence result it is useful to express the sufficient horizon
in dimensionless form:
\[
 \kappa\overline T_H
 =2z+e^z\left(\frac{2z}{3\pi}+\frac{3\pi}{2z}\right),
 \qquad z=\kappa H.
\]
Figure~\ref{fig:divergence-bound} plots this sufficient bound.
It is large for very small or very large delays. This shape concerns
the bound supplied by our chosen negative direction; it does not
identify the first horizon at which the projected rule diverges.

All formula-based plots use the exact expressions in
Theorems~\ref{prop:finite} and \ref{prop:naive}. The numerical tables can
be reproduced directly from these expressions.

\FloatBarrier
\clearpage
\section{Proof of Theorem~\ref{prop:finite}}
\label{sec:exact-proof}

We first compute the value, then derive the strategy. Set
\[
 X_t=\frac{r_t-x}{\sigma},\qquad m=\frac{\theta-x}{\sigma},
 \qquad \eta_t=\sigma\gamma_t,
 \qquad A_t=\kappa[1+\kappa(T-t)].
\]
Then $dX_t=\kappa(m-X_t)dt+dW_t$, $X_0=0$, and
$\int_{0}^T\gamma_t\,dr_t=\int_{0}^T\eta_t\,dX_t$.

\subsection{The Gaussian optimization problem}

On every finite horizon, the law of $X$ is equivalent to Wiener measure
$\Ww$ by \citet[Theorem~2]{Hitsuda1968}. Indeed,
\[
 X_t=m(1-e^{-\kappa t})+W_t
 -\int_0^t\int_0^u\kappa e^{-\kappa(u-s)}\,dW_s\,du,
\]
whose Volterra kernel is square-integrable and whose mean has
square-integrable derivative. Its Cameron--Martin shift is therefore
admissible on each finite horizon. The inverse relation
$dW_t=dX_t-\kappa(m-X_t)dt$ gives the likelihood
\begin{align}
 \ell_T:=\log\frac{d\Pp}{d\Ww}
 &=\int_0^T\kappa(m-X_t)\,dX_t
 -\frac{\kappa^2}{2}\int_0^T(m-X_t)^2dt \notag\\
 &=c+\int_0^T a(t)\,dX_t
 +\int_0^T\int_0^t f(t,s)\,dX_s\,dX_t,
 \label{eq:likelihood}
\end{align}
where
\[
 a(t)=A_tm,\qquad
 c=-\frac{\kappa^2m^2T}{2}-\frac{\kappa^2T^2}{4},\qquad
 f(t,s)=-\kappa-\kappa^2(T-\max(t,s)).
\]
For clarity, It\^o's formula gives
\[
 \ell_T=\frac{\kappa T}{2}-\frac{\kappa^2m^2T}{2}
 +\kappa mX_T+\kappa^2m\int_0^T X_tdt
 -\frac{\kappa}{2}X_T^2-\frac{\kappa^2}{2}\int_0^T X_t^2dt.
\]
Using $X_t^2=t+2\int_0^tX_s\,dX_s$ under $\Ww$ and applying
Fubini identifies $c,a,f$ above. These stochastic identities also hold
under the equivalent physical law. 

We apply \citet[Theorem~1]{Dolinsky2025} on canonical path space, with
the Ornstein--Uhlenbeck law of $X$ as the physical measure and $\Ww$ as
the reference measure. Normalization preserves the price filtration and
maps $\gamma$ to $\eta=\sigma\gamma$ bijectively. Thus our problem is
exactly the exponential-loss minimization problem in that theorem. Its
inputs are the likelihood coefficients $c,a,f$ in
\eqref{eq:likelihood} and the observation function
$\tau(t)=(t-H)^+$. For $s>0$, the inverse observation time is
$b_s=\min(s+H,T)$: the price increment at time $s$ becomes observable
at time $b_s$, or remains unobserved before the horizon.

The theorem gives attainment over the full class $\A_H(T)$ and determines
two kernels: a Volterra trading kernel $K$, extended by zero when
$s>t$, and a symmetric Gaussian kernel. We denote the lower-triangular
restriction of the latter by $g$; only these values enter our formulas.
In the Brownian representation, $g$ means its Volterra extension, zero
for $s>t$. The likelihood kernel $f$ retains its symmetric definition
on $[0,T]^2$. We use $K$ for the kernel denoted by $\kappa$ in that theorem, to
distinguish it from the mean-reversion speed. Their defining relation is
\begin{equation}
 f(t,s)-K(t,s)+g(t,s)
 =\int_s^T[f(t,u)-K(t,u)]g(u,s)du,
 \qquad 0\le s\le t\le T.
 \label{eq:joint-kernels}
\end{equation}
On the lower triangle, the support restrictions are $K(t,s)=0$ for
$t<b_s$ and $g(t,s)=0$ for $t>b_s$, with endpoint values immaterial for the integrals. For
$s\le t<b_s$, \eqref{eq:joint-kernels} therefore reduces to
\begin{equation}
 g(t,s)+f(t,s)=\int_s^{b_s} f(t,u)g(u,s)du,
 \qquad s\le t\le b_s,
 \label{eq:g-equation}
\end{equation}
where we use the continuous representative on $[s,b_s]$. Once $g$ is
known, the same relation determines $K$ on the observed region.
Indeed, for $s\le t<b_s$ and $s\le u\le b_s$, the inequality
$b_u\ge b_s>t$ implies $K(t,u)=0$; this justifies the reduction
to \eqref{eq:g-equation}.
All identities below involving kernels are understood almost everywhere;
continuous representatives on the indicated triangles are used for
boundary calculations. The theorem gives the value formula
\begin{equation}
 \log V_H(T)=c-\int_0^T\int_0^t f(t,s)\widetilde g(t,s)dsdt
 -\frac12\int_0^T\int_0^t g(t,s)^2dsdt,
 \label{eq:gaussian-value}
\end{equation}
where $\widetilde g(t,s)=g(t,s)-\int_0^s g(t,u)g(s,u)du$.
Thus the cited theorem supplies global optimality and attainment. Our
remaining task is to solve for $g$, evaluate this value, and express the
resulting trading kernel through observed prices.

\subsection{The kernel and the value}

Fix $s<T$ and put
\[
 h_s=b_s-s,\qquad q_s=T-b_s,\qquad z_s=1+\kappa q_s,
 \qquad D_s=z_s\sinh(\kappa h_s)+\cosh(\kappa h_s).
\]
For fixed $s$, split the integral in \eqref{eq:g-equation} at $u=t$.
On $u<t$ the derivative $\partial_tf(t,u)=\kappa^2$, while on $u>t$
it is zero. The two boundary terms cancel because $f$ is continuous.
Thus $\partial_tg(t,s)+\kappa^2=\kappa^2\int_s^t g(u,s)du$. At $t=s$
this gives the left boundary condition; evaluating the original equation
at $t=b_s$, where $f(b_s,u)=-\kappa(1+\kappa q_s)$, gives the right one. Hence
$y(t)=g(t,s)$ satisfies
\[
 y''=\kappa^2y,\qquad y'(s)=-\kappa^2,
 \qquad y(b_s)+(q_s+\kappa^{-1})y'(b_s)=0.
\]
Conversely, these conditions imply \eqref{eq:g-equation}. Indeed,
$y'(t)=-\kappa^2+\kappa^2\int_s^t y(u)du$, so the residual in that
integral equation has zero derivative. At $b_s$ its value is
$y(b_s)+(q_s+\kappa^{-1})y'(b_s)=0$. Solving the boundary-value
problem yields
\begin{equation}
 g(t,s)=\kappa\,
 \frac{z_s\cosh(\kappa(b_s-t))+\sinh(\kappa(b_s-t))}{D_s},
 \qquad s\le t\le b_s.
 \label{eq:g-explicit}
\end{equation}
Since $D_s\ge1$, the solution is unique and bounded on finite horizons.
In particular,
\begin{equation}
 g(s,s)-\kappa=
 \begin{cases}
 \displaystyle\frac{\kappa^2e^{-2\kappa H}(T-H-s)}
 {1+\beta_H(T-H-s)},&s<(T-H)^+,\\[2mm]
 0,&s\ge(T-H)^+.
 \end{cases}
 \label{eq:g-diagonal}
\end{equation}

For brevity, write
\[
 I_1=\int_0^T\int_0^t fg\,dsdt,\qquad
 I_2=\int_0^T\int_0^t g^2\,dsdt,\qquad
 I_3=\int_0^T\int_0^t f(t,s)\int_0^s g(t,u)g(s,u)dudsdt.
\]
The kernels are bounded on the finite integration regions, so Fubini
applies. For each fixed $u$, the product $g(t,u)g(s,u)$ is supported
on $[u,b_u]^2$. Symmetry of $f$ changes the triangular integral in
$(t,s)$ into one half of the integral over that square. Hence
\begin{align*}
 I_3&=\frac12\int_0^T\int_u^{b_u}g(t,u)
 \left(\int_u^{b_u}f(t,s)g(s,u)ds\right)dtdu\\
 &=\frac12(I_1+I_2).
\end{align*}
Consequently, \eqref{eq:gaussian-value} reduces to
$\log V_H(T)=c-I_1/2$. Evaluating \eqref{eq:g-equation} on the diagonal
and integrating gives
\[
 I_1=\int_0^T\int_s^{b_s}f(s,t)g(t,s)\,dtds
 =\int_0^T[g(s,s)+f(s,s)]ds
 =\int_0^Tg(s,s)ds-\kappa T-\frac{\kappa^2T^2}{2}.
\]
The quadratic terms cancel, and therefore
\begin{equation*}
 \log V_H(T)=-\frac{\kappa^2m^2T}{2}
 -\frac12\int_0^T[g(s,s)-\kappa]ds.
\end{equation*}
With $L=(T-H)^+$, change variables $q=L-s$ in
\eqref{eq:g-diagonal}. Since
$\kappa^2e^{-2\kappa H}/(2\beta_H)=\lambda_H$, this gives
\begin{align*}
 \CE_H(T)&=\nu T+\frac{\kappa^2e^{-2\kappa H}}{2}
 \int_0^L\frac{q}{1+\beta_Hq}dq\\
 &=\nu T+\lambda_H\left[L-\frac{\log(1+\beta_HL)}{\beta_H}\right],
\end{align*}
which proves \eqref{eq:exact-ce}.

For fixed $H>0$,
\[
 \log(1+\beta_H(T-H))=\log T+\log\beta_H+O(T^{-1}).
\]
Substituting in \eqref{eq:exact-ce} gives \eqref{eq:long-ce}.

\subsection{The optimal strategy}
\label{sec:portfolio}

If $H\ge T$, the trading kernel vanishes and the stated strategy is
deterministic. Hence suppose $H<T$. With $g$ obtained above, equation~(9) of \citet{Dolinsky2025} gives the optimal normalized strategy as
\begin{equation*}
 \widehat\eta_t=a(t)+\int_0^{(t-H)^+}K(t,s)\,dX_s.
\end{equation*}
For $t>H$, let $d=t-H$. On $s\le d$, we have $g(t,s)=0$, so
\eqref{eq:joint-kernels} gives
\[
 f(t,s)-K(t,s)=\int_s^{s+H}[f(t,u)-K(t,u)]g(u,s)du,
 \qquad 0\le s\le d,
\]
where $K(t,u)=0$ for $u>d$. On this integration interval $f(t,u)=-A_t$.
Throughout this calculation $g(u,s)=0$ when $u>s+H$.
Also, \eqref{eq:g-explicit} gives
\[
 \int_s^{s+H}g(u,s)du=1-D_s^{-1},\qquad
 D_s=e^{\kappa H}+\kappa(T-H-s)\sinh(\kappa H).
\]
Thus
\begin{equation}
 K(t,s)-\int_s^d K(t,u)g(u,s)du=-A_tD_s^{-1}.
 \label{eq:K-resolvent}
\end{equation}

The proof of \citet[Theorem~1]{Dolinsky2025} constructs an equivalent
Gaussian measure with a Brownian motion $B$ such that
\[
 dX_u=dB_u-\left(\int_0^u g(u,s)dB_s\right)du.
\]
Put $M_u=\int_0^u D_s^{-1}dB_s$, $u\le T-H$. The deterministic kernels are square-integrable on the finite horizon,
so stochastic Fubini under this Gaussian measure and
\eqref{eq:K-resolvent} imply
\[
 \int_0^dK(t,s)dX_s
 =\int_0^d\left[K(t,s)-\int_s^dK(t,u)g(u,s)du\right]dB_s
 =-A_tM_d.
\]
It follows that $\widehat\eta_t=A_tm-A_tM_{t-H}$ for $t>H$;
for $t\le H$ it equals $A_tm$.

To express $M$ in terms of observed prices, integrate the representation
of $X$ and use \eqref{eq:g-explicit}. For $u\le T-H$,
\[
 X_u=\int_0^uJ(u,s)dM_s,
\]
where $J(u,s)=1$ for $s<u-H$, while for $s\ge(u-H)^+$,
\[
 J(u,s)=z_s\sinh(\kappa(H-u+s))+\cosh(\kappa(H-u+s)).
\]
The two expressions agree at $s=u-H$, and
$J(u,u)=D_u$, $\partial_sJ(u,s)=\kappa z_s\cosh(\kappa(H-u+s))$
on the second interval. Integration by parts gives
\begin{equation}
 X_u=D_uM_u-\int_{(u-H)^+}^u
 \kappa z_s\cosh(\kappa(H-u+s))M_sds.
 \label{eq:observed-M}
\end{equation}
Set $\mathcal Y_t=A_tM_{t-H}$ for $t>H$ and $\mathcal Y_t=0$ otherwise.
Since $A_{s+H}=\kappa z_s$ and $A_t/D_{t-H}=C_t$,
\eqref{eq:observed-M} becomes
\[
 \mathcal Y_t=C_t\left[X_{t-H}+\int_{t-H}^t
 \cosh(\kappa(H-t+s))\mathcal Y_sds\right].
\]
Taking $Y=\sigma\mathcal Y$ proves \eqref{eq:volterra}, and
$\widehat\eta_t=A_tm-\mathcal Y_t$ proves
\eqref{eq:optimal-portfolio}. Equivalence of the measures transfers these identities to $\Pp$.

Existence and uniqueness of the admissible optimizer follow from
\citet[Theorem~1]{Dolinsky2025}. The preceding identities express that
optimizer through the continuous, $\G$-adapted solution of
\eqref{eq:volterra}, establishing existence of $Y$. Since the kernel is
bounded on the finite horizon, Gronwall's inequality gives uniqueness
of this solution. The optimal strategy is unique up to
$dt\otimes\Pp$-null sets by the cited theorem.
This completes the proof of Theorem~\ref{prop:finite}. \qed

\section{Proof of Theorem~\ref{prop:naive}}
\label{sec:naive}

The Ornstein--Uhlenbeck conditional mean is
$\E[\theta-r_t\mid\F_s]=e^{-\kappa(t-s)}(\theta-r_s)$ for $s\le t$,
which gives \eqref{eq:naive-formula-main}. This strategy is continuous
and $\G$-adapted. Since
\[
 \E(\theta-r_s)^2=(\theta-x)^2e^{-2\kappa s}
 +\frac{\sigma^2}{2\kappa}(1-e^{-2\kappa s}),
\]
its squared position has a finite time integral in expectation, proving
both admissibility claims.

For the divergence claim, put
\[
 m=\frac{\theta-x}{\sigma},\qquad
 A_t=\kappa[1+\kappa(T-t)],\qquad d_t=me^{-\kappa t},
\]
and define the centered Gaussian processes
\[
 U_t=\int_0^t e^{-\kappa(t-s)}dW_s,\qquad
 V_t=e^{-\kappa H}U_{(t-H)^+}.
\]
Then $(\theta-r_t)/\sigma=d_t-U_t$, and the gain is
\begin{equation}
 G_T^N=\int_0^T A_t(d_t-V_t)
 \{\kappa(d_t-U_t)dt+dW_t\}.
 \label{eq:naive-gain}
\end{equation}

We isolate one Gaussian direction. For a deterministic unit vector
$e\in L^2[0,T]$, write
\[
 N=\int_0^T e(s)dW_s,\qquad
 W_t=N\int_0^t e(s)ds+W_t^\perp.
\]
Here $N$ is standard normal and independent of the Gaussian process
$W^\perp$. The gain in \eqref{eq:naive-gain} is an $L^2$ limit of
polynomials of degree at most two in Gaussian increments. Substituting
this decomposition in those polynomials and passing to the limit gives
\[
 G_T^N=R+LN+q(e)N^2,
\]
where $R,L$ are finite random variables measurable with respect to
$W^\perp$. To compute the deterministic quadratic coefficient, set
$z_e(t)=\int_0^t e^{-\kappa(t-s)}e(s)ds$, so
$z_e'=e-\kappa z_e$. The terms quadratic in $N$ give
\begin{align*}
 q(e)&=e^{-\kappa H}\int_H^T
 A_tz_e(t-H)[\kappa z_e(t)-e(t)]dt\\
 &=-e^{-\kappa H}\int_H^T A_tz_e(t-H)z_e'(t)dt.
\end{align*}
If $q(e)\le-1/2$, conditioning on $W^\perp$ yields
\[
 \E[e^{-G_T^N}\mid W^\perp]
 =\frac{e^{-R}}{\sqrt{2\pi}}\int_{\R}
 e^{-(q(e)+1/2)y^2-Ly}dy=\infty.
\]
This also holds at equality, for every finite $L$.

It remains to choose $e$. Suppose $T\ge2H$, put
$\omega=3\pi/(2H)$, and take
\[
 z(t)=\begin{cases}\sin(\omega t),&0\le t\le2H,\\
 0,&2H<t\le T,\end{cases}
 \qquad h=z'+\kappa z,\qquad e=\frac{h}{\|h\|_2}.
\]
Since $z(0)=z(2H)=0$, this function is absolutely continuous,
$z_e=z/\|h\|_2$, and
$\|h\|_2^2=H(\kappa^2+\omega^2)$. On $[H,2H]$,
$z(t-H)=\cos(\omega t)$ and $z'(t)=\omega\cos(\omega t)$.
Consequently,
\[
 q(e)=-\frac{e^{-\kappa H}\omega}{H(\kappa^2+\omega^2)}
 \int_H^{2H}A_t\cos^2(\omega t)dt
 \le-\frac{e^{-\kappa H}\omega\kappa^2(T-2H)}
 {2(\kappa^2+\omega^2)},
\]
where we used $A_t\ge\kappa^2(T-2H)$ and
$\int_H^{2H}\cos^2(\omega t)dt=H/2$. Thus $q(e)\le-1/2$ whenever
\[
 T\ge2H+\frac{e^{\kappa H}(\kappa^2+\omega^2)}{\omega\kappa^2}
 =2H+e^{\kappa H}\left(\frac{2H}{3\pi}
 +\frac{3\pi}{2\kappa^2H}\right)=\overline T_H.
\]
The conditional Gaussian integral above now proves
\eqref{eq:naive-infinite-main} for all $x,\theta$ and $\sigma>0$.
\qed


\begin{thebibliography}{99}

\bibitem[Bank \protect\APAand{} Dolinsky(2021)]{BankDolinsky2021}
Bank, P., \& Dolinsky, Y. (2021).
Short communication: A note on utility indifference pricing with delayed information.
\emph{SIAM Journal on Financial Mathematics, 12}, SC31--SC43.
\url{https://doi.org/10.1137/20M1379630}

\bibitem[Bj\"ork et al.(2010)Bj\"ork, Davis, \protect\APAand{} Land\'en]{BjorkDavisLanden2010}
Bj\"ork, T., Davis, M. H. A., \& Land\'en, C. (2010).
Optimal investment under partial information.
\emph{Mathematical Methods of Operations Research, 71}, 371--399.
\url{https://doi.org/10.1007/s00186-010-0301-x}

\bibitem[Dolinsky(2025)]{Dolinsky2025}
Dolinsky, Y. (2025).
Exponential utility maximization with delay in a continuous time Gaussian framework.
\emph{Systems \& Control Letters, 206}, 106270.
\url{https://doi.org/10.1016/j.sysconle.2025.106270}

\bibitem[Dolinsky \protect\APAand{} Zouari(2020)]{DolinskyZouari2020}
Dolinsky, Y., \& Zouari, J. (2020).
Market delay and $G$-expectations.
\emph{Stochastic Processes and their Applications, 130}, 694--707.
\url{https://doi.org/10.1016/j.spa.2019.03.007}

\bibitem[Dolinsky \protect\APAand{} Zuk(2023)]{DolinskyZuk2023}
Dolinsky, Y., \& Zuk, O. (2023).
Short communication: Exponential utility maximization in a discrete time Gaussian framework.
\emph{SIAM Journal on Financial Mathematics, 14}, SC31--SC41.
\url{https://doi.org/10.1137/23M1576074}

\bibitem[Dolinsky \protect\APAand{} Zuk(2025)]{DolinskyZuk2025}
Dolinsky, Y., \& Zuk, O. (2025).
Explicit computations for delayed semistatic hedging.
\emph{SIAM Journal on Financial Mathematics, 16}, 29--52.
\url{https://doi.org/10.1137/23M1626888}

\bibitem[Guasoni et al.(2021)Guasoni, Nagy, \protect\APAand{} R\'asonyi]{GNR2021}
Guasoni, P., Nagy, L., \& R\'asonyi, M. (2021).
Young, timid, and risk takers.
\emph{Mathematical Finance, 31}, 1332--1356.
\url{https://doi.org/10.1111/mafi.12329}

\bibitem[Hitsuda(1968)]{Hitsuda1968}
Hitsuda, M. (1968).
Representation of Gaussian processes equivalent to Wiener process.
\emph{Osaka Journal of Mathematics, 5}, 299--312.
\url{https://doi.org/10.18910/4116}

\bibitem[Lakner(1995)]{Lakner1995}
Lakner, P. (1995).
Utility maximization with partial information.
\emph{Stochastic Processes and their Applications, 56}, 247--273.
\url{https://doi.org/10.1016/0304-4149(94)00073-3}

\bibitem[Lakner(1998)]{Lakner1998}
Lakner, P. (1998).
Optimal trading strategy for an investor: The case of partial information.
\emph{Stochastic Processes and their Applications, 76}, 77--97.
\url{https://doi.org/10.1016/S0304-4149(98)00032-5}

\bibitem[Mania \protect\APAand{} Santacroce(2010)]{ManiaSantacroce2010}
Mania, M., \& Santacroce, M. (2010).
Exponential utility maximization under partial information.
\emph{Finance and Stochastics, 14}, 419--448.
\url{https://doi.org/10.1007/s00780-009-0114-z}

\bibitem[Nourdin \protect\APAand{} Poly(2012)]{NourdinPoly2012}
Nourdin, I., \& Poly, G. (2012).
Convergence in law in the second Wiener/Wigner chaos.
\emph{Electronic Communications in Probability, 17}, 1--12.
\url{https://doi.org/10.1214/ECP.v17-2023}



\bibitem[Saporito \protect\APAand{} Zhang(2019)]{SaporitoZhang2019}
Saporito, Y. F., \& Zhang, J. (2019).
Stochastic control with delayed information and related nonlinear master equation.
\emph{SIAM Journal on Control and Optimization, 57}, 693--717.
\url{https://doi.org/10.1137/17M1154746}

\end{thebibliography}
\end{document}